\documentclass[prd,twocolumn]{revtex4}
\usepackage{graphicx, epsfig}
\usepackage{color}
\usepackage{mathrsfs}
\usepackage{bm}
\usepackage{amsmath,amssymb}% for \eqref
\usepackage[caption=false]{subfig}

\newcommand{\be}{\begin{equation}}
\newcommand{\ee}{\end{equation}}
\newcommand{\ba}{\begin{eqnarray}}
\newcommand{\ea}{\end{eqnarray}}

\newcommand{\gapp}{\mathrel{\raise.3ex\hbox{$>$}\mkern-14mu
              \lower0.6ex\hbox{$\sim$}}}
\newcommand{\lapp}{\mathrel{\raise.3ex\hbox{$<$}\mkern-14mu
              \lower0.6ex\hbox{$\sim$}}}

\begin{document}
\title{Monopole-antimonopole Interaction Potential}
\author{Ayush Saurabh$^\dag$, Tanmay Vachaspati$^{\dag *}$}
\affiliation{
$^\dag$Physics Department, Arizona State University, Tempe, AZ 85287, USA. \\
$^*$Maryland Center for Fundamental Physics, University of Maryland,
                    College Park, MD 20742, USA.
}

%%%%%%%%%%%%%%%%%%%%%%%%%%%%%%%%%%%%%%%%%%%%%%%%%%%%%%%
\begin{abstract}
\noindent
We numerically study the interactions of twisted monopole-antimonopole pairs in the 't Hooft-Polyakov model 
for a range of values of the scalar to vector mass ratio. We also recover the sphaleron solution at maximum twist
discovered by Taubes~\cite{taubes}, and map out its energy and size as functions of parameters.
\end{abstract}
%%%%%%%%%%%%%%%%%%%%%%%%%%%%%%%%%%%%%%%%%%%%%%%%%%
\maketitle

%\section{Introduction}

Magnetic monopoles are novel solutions in a large class of non-Abelian gauge theories~\cite{thooft,Polyakov:1974ek}.
They are also an essential prediction of all Grand Unified Theories of particle physics. They have been studied
for their unconventional classical and quantum properties~\cite{manton}, and experiments are currently underway
to find cosmological monopoles~\cite{icecube,antares} as well as in particle accelerators~\cite{moedal}. 

In spite of the long history of monopoles, there are certain questions that have not been fully resolved. Key
among these is to discover particle physics processes that can create magnetic monopoles~\cite{tanmaycreation}.
Dynamics that involves both monopoles and antimonopoles, has not received much attention~\cite{tanmayscattering}.
On the other hand, monopole-{\it monopole} dynamics has been beautifully resolved in the moduli 
approximation~\cite{mantonremarks}. % as well as numerically~\TV{cite}. 

An important feature in the monopole-antimonopole system is that the monopole and antimonopole can have a 
relative twist (see Sec.~\ref{mmbarconfiguration}). This additional degree of freedom has profound consequences 
for the interaction energy of a monopole and antimonopole. In particular it enables the existence of static bound state 
solutions, now known as a ``sphaleron'', as first argued by 
Taubes~\cite{taubes}. The sphaleron was rigorously shown to exist in the 
special case of vanishing scalar mass by Taubes \cite{taubes,taubessecondpart} 
and for non-vanishing scalar mass in \cite{groisserthesis}.  The Morse theory 
analysis used by Taubes in an $SU(2)$ model was used by Manton for the 
physically
relevant electroweak theory~\cite{mantonelectroweak}. This resulted in the discovery of the ``electroweak sphaleron'' that
interpolates between vacua of different Chern-Simons number and is critical to understanding the violation of baryon 
number in electroweak theory. 

Based on a qualitative understanding of the scalar and vector forces between a monopole and an antimonopole
at separation $d$, Taubes sketched the interaction potential as
\be 
V(d,\gamma) = 4\pi\left(-\frac{1}{d} -\frac{2e^{-d}}{d} \cos \gamma 
-\frac{e^{-\sqrt{\lambda} d}}{d} (1-e^{-d})\right)
\label{taubesequation}
\ee
where the first term on the right hand side is the usual attractive Coulomb interaction, the second term is 
a correction term which represents short range interactions mediated by the two massive vector bosons 
$W^{\pm}$, $\gamma$ is the relative twist angle, and the last term is due to scalar interactions.
(Note: in Taubes' notation, the twist is called $\theta$ where $\theta = \pi-\gamma$.) 
This vector interaction is attractive for $\cos\gamma > 0$ and repulsive for $\cos\gamma < 0$, in which case
the attractive Coulomb and repulsive forces can balance at some separation, leading to a saddle point solution.
Any perturbation to this solution that untwists the pair will destabilize the solution, and the monopole and antimonopole
will eventually radiate, as seen in \cite{tanmayscattering}.

We will see that the expression for $V(d,\gamma)$ in Eq.~(\ref{taubesequation}) 
provides a good
qualitative picture but does not provide a good fit to the numerical data. This can be expected
because the terms in Eq.~(\ref{taubesequation}) assume point-like monopoles. In reality,
monopoles are extended objects and a monopole-antimonopole can partially annihilate as
they are brought closer together, {\it i.e.} when the cores of the 
monopole-antimonopole overlap there is a reduction
in the volume occupied by the cores. Further, the reduction of energy depends 
on the extent of partial annihilation that, in turn, can depend on the amount 
of twist. Thus the actual potential
can be more complicated than that given by Eq.~(\ref{taubesequation}).

A goal of our work is to rigorously determine $V(d, \gamma)$. Our numerical 
approach can be applied to {\it any} values of model parameters, and we are 
able to reconstruct all the fields for the monopole-antimonopole system. In 
particular, we calculate their interaction energy, the size, and energy, of the 
monopole-antimonopole bound state, for a range of couplings. For a special 
twist and separation we can recover the sphaleron that was also investigated 
numerically in \cite{Kleihaus:1999sx} by solving the static equations of motion 
by first taking an
axially symmetric ansatz for the fields. In contrast, we employ constrained relaxation over an entire three dimensional 
grid without assuming any symmetries, and we also study monopole-antimonopole pairs away from the sphaleron.

We start by introducing the model and magnetic monopoles in Sec.~\ref{su2monopoles}, and the
monopole-antimonopole configuration in Sec.~\ref{mmbarconfiguration}. We describe our choice of
the ``twisted dipole'' gauge in Sec.~\ref{twisteddipolegauge}, which is crucial to the success of our 
numerical scheme described in Sec.~\ref{numerics}. Our results and conclusion can be found in
Sec.~\ref{results}.

\section{$SU(2)$ Monopoles}
\label{su2monopoles}

We consider 't Hooft-Polyakov monopoles~\cite{thooft,Polyakov:1974ek}
in the $SU(2)$ model
\be
\mathcal{L}=\frac{1}{2}(D^{\mu}\phi)^{a}(D_{\mu}\phi)^{a}-\frac{1}{4}W^{a\mu\nu}W_{\mu\nu}^{a}
-\frac{\lambda}{4}(\phi^{a}\phi^{a}-\eta^{2})^{2}
\ee
where $a=1,2,3$, the covariant derivative is defined as,
\be
(D_{\mu}\phi)^{a}=\partial_{\mu}\phi^{a}+g\epsilon^{abc}W_{\mu}^{b}\phi^{c}
\ee
and the gauge field strength is given as 
\be
W_{\mu\nu}^{a}=\partial_{\mu}W_{\nu}^{a}-\partial_{\nu}W_{\mu}^{a}+g\epsilon^{abc}W_{\mu}^{b}W_{\nu}^{c}
\ee

The equations of motion are written as~\cite{tanmaycreation},
\ba
\partial_t^2 \phi^a &=& \nabla^2 \phi^a
- g \epsilon^{abc}\partial_i\phi^b W_i^c- g \epsilon^{abc} (D_i \phi)^b W_i^c \nonumber \label{eqphi} \\ 
&&
- \lambda (\phi^b\phi^b-\eta^2)\phi^a- g \epsilon^{abc} \phi^b \Gamma^c \\
\partial_t W^a_{0i} &=& \nabla^2 W^a_i
+ g \epsilon^{abc} W^b_j \partial_j W^c_i - g\epsilon^{abc} W^b_j W^c_{ij} \nonumber \label{eqgauge} \\
&&
- D_i \Gamma^a - g\epsilon^{abc} \phi^b (D_i\phi)^c \\
\partial_t \Gamma^a &=& \partial_i W^a_{0i}
- g_p^2 [ \partial_i (W^a_{0i})  +g\epsilon^{abc}W^b_i W^c_{0i} \nonumber \\
&&
+ g\epsilon^{abc}\phi^b (D_t \phi)^c ]
\label{Gammaeq}
\ea
where we are using temporal gauge, $W_{0}^{a}=0$, $\Gamma^{a}=\partial_{i}W_{i}^{a}$
are introduced as new variables, and $g_p^2$ is a numerical parameter that we can choose
to ensure numerical stability. 
By rescaling the fields and spatial coordinates appropriately, and setting the vacuum 
expectation value and coupling constants to one, that is, $\eta=g=1$, it is easily seen that 
$\lambda$ is the only parameter in the theory that controls the mass and size of the monopoles. 

Varying the action with respect to the metric gives us the following
expression for energy of a given static configuration,
\begin{align}
E=\int d^{3}x[\frac{1}{2}(D_{i}\phi)^{a}(D_{i}\phi)^{a}&+\frac{1}{4}W_{ij}^{a} W_{ij}^{a} \nonumber \\ 
&+\frac{\lambda}{4}(\phi^{a}\phi^{a}-1^{2})^{2}]
\label{energydensity}
\end{align}

Our goal in the present analysis is to solve for static monopole-antimonopole
configurations that minimize the above energy functional, with the constraints
that fix the locations and relative twist of the monopoles. An essential
condition for the existence of finite energy solutions is that the
terms in the integrand vanish individually at spatial infinity. This
requires 
\begin{align}
	\phi^{a}\phi^{a} \rightarrow 1, \ \ (D_{i}\phi)^{a} \rightarrow 0, \ \
W_{ij}^{a}&\rightarrow 0
\label{asymptotics}
\end{align}
at spatial infinity.

A non-zero vacuum expectation value of $\phi^a$ spontaneously breaks the $SU(2)$
symmetry to a $U(1)$ subgroup and two of the three gauge fields acquire a mass
while the third is the ``photon''.
This electromagnetic gauge field can be expressed as
\be
A_{\mu}=\hat{\phi}^{a}W_{\mu}^{a}.
\label{emA}
\ee
The electromagnetic field strength is now defined as
\ba
F_{\mu\nu} &=&
\hat{\phi}^a W^a_{\mu\nu} - \epsilon^{abc} \hat{\phi}^a (D_\mu \hat{\phi})^b (D_\nu \hat{\phi})^c \\
&=& 
\partial_\mu A_\nu - \partial_\nu A_\mu 
- \epsilon^{abc} \hat{\phi}^a \partial_\mu \hat{\phi}^b \partial_\nu \hat{\phi}^c, \ \ {\rm if}\ |\phi|=1. \nonumber
\label{emF}
\ea
These expressions are identical to those proposed in~\cite{thooft} in the region outside
the monopole. Also note that we have set $g=1=\eta$ in these expressions.

\section{Monopole-antimonopole Configuration}
\label{mmbarconfiguration}

We are going to solve the equations of motion presented in the previous
section numerically using a fixed point iteration scheme. This scheme
will relax an initial guess field configuration at each iteration step. 
As with all relaxation schemes, a good initial guess is important for our method to converge.

To understand the ansatz that we used for our analysis, we start with
the field configuration of a spherically symmetric magnetic monopole. We choose the Higgs isovector such that it always points along the radial position vector, that is, $\hat{\phi}^{a}=\hat{r}^{a}$, where $\hat{r}^{a}=r^{a}/|\vec{r}|$. This means that we can write
our Higgs fields as
\be
\phi^{a}=h(r)\hat{r^{a}}
\ee
The direction for gauge fields can be shown, by satisfying the condition
that the covariant derivative of the Higgs fields vanish at spatial infinity, to take the form below
\be
W_{i}^{a}=\frac{(1-k(r))}{r}\epsilon^{aij}\hat{r}^{j}
\ee
To solve for the profile functions $h(r)$ and $k(r)$, we plug these
last expressions into the general equations of motions. This gives
us two coupled ordinary differential equations as follows,
\begin{align}
h''(r)+\frac{2}{r} h'(r) &= \frac{2}{r^2} k(r)^2 h(r) -\lambda \left(h(r)^2-1 \right) h(r)  \\
k''(r)&=\frac{1}{r^2}(k(r)^2-1)k(r)+h(r)^2 k(r)
\end{align}

\begin{figure}[h]
  \includegraphics[height=0.25\textwidth,angle=0]{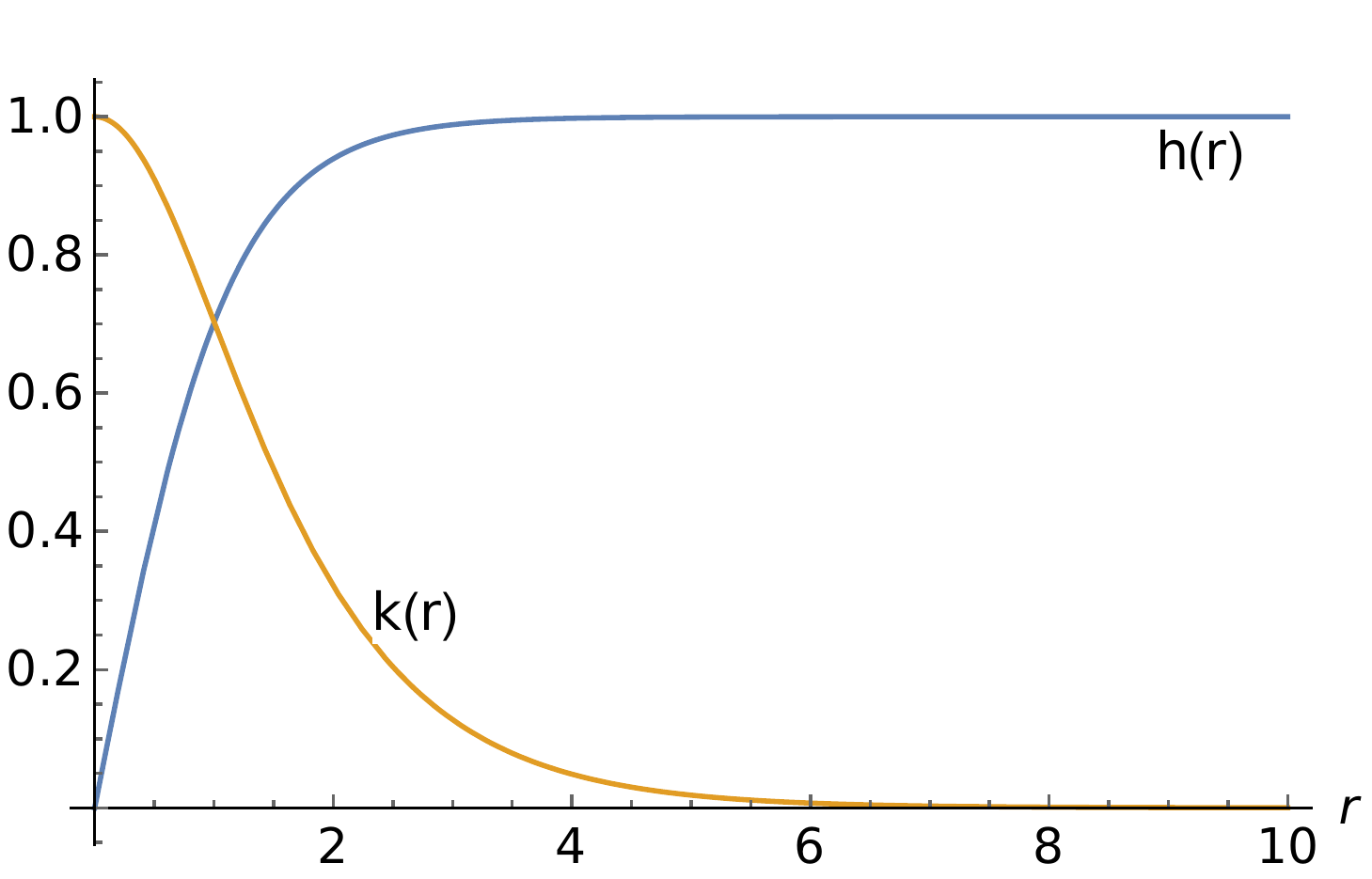}
  \caption{Numerically generated profile functions, $h(r)$ and $k(r)$, for $\lambda=1$.
 }
 \label{hk}
\end{figure}

These differential equations in one dimension are solved numerically with the
Gauss-Newton method for different values of $\lambda$ and with boundary
conditions, $h(r)\rightarrow1$ and $k(r)\rightarrow0$ as $r\rightarrow\infty$,
and $h(r)\rightarrow0$ and $k(r)\rightarrow1$ as $r\rightarrow0$.
Fig.~\ref{hk} shows a plot of these profile functions for $\lambda=1$.
The mass of the monopole is shown in Table~\ref{masstable} for sample values 
of $\lambda$. We will use these solutions in our initial guess
for the monopole-antimonopole field configuration.

\begin{table}[ht]
\caption{Mass of the monopole as a function of $\lambda$}
\begin{tabular}{c | c}
$\lambda$ & Mass in units of $4\pi$ \\
\hline
0.0 & 1.000 \\
0.25 & 1.185 \\
0.50 & 1.232 \\
0.75 & 1.264 \\
1.0 & 1.287 
\end{tabular}
\label{masstable}
\end{table}

\begin{figure}
	\includegraphics[trim={1.4cm 0 0 0},width=0.23\textwidth,angle=0]{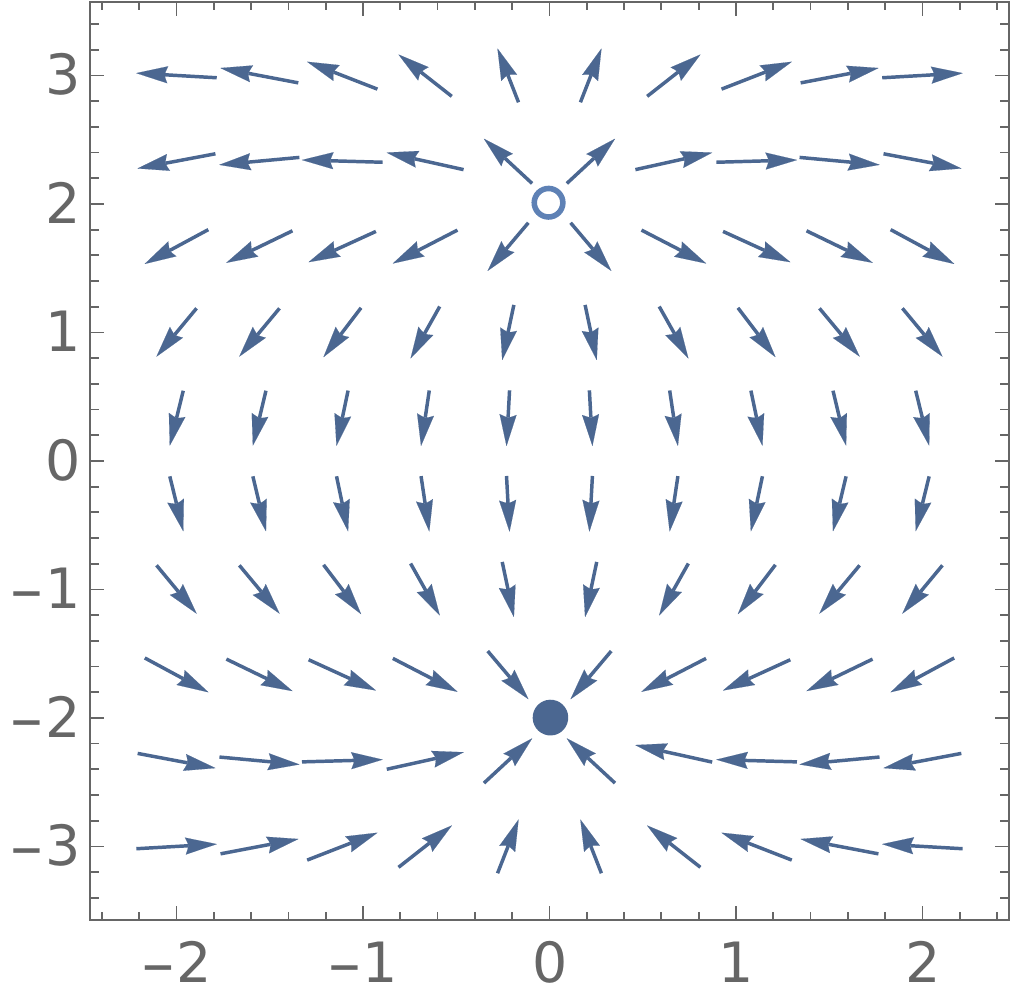}
	\includegraphics[trim={0.5cm 0 0.4cm 0},width=0.23\textwidth,angle=0]{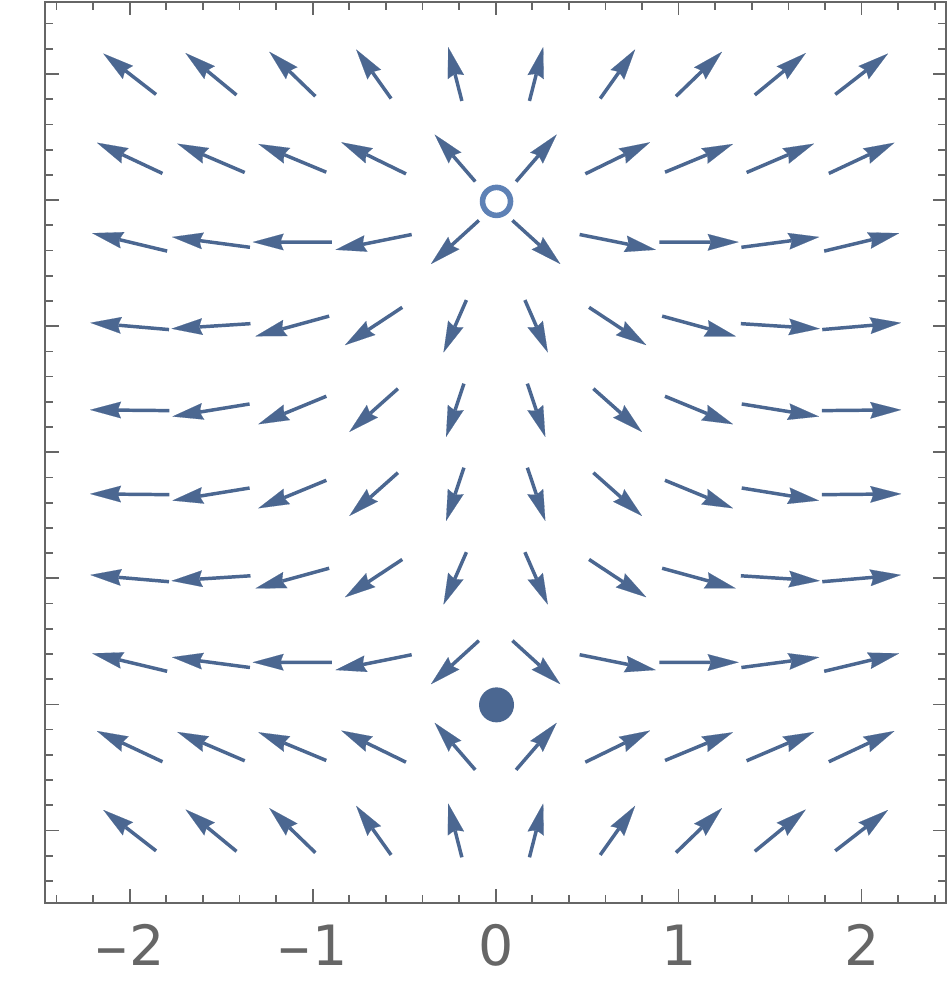}
\caption{Higgs vectors in the $xz-$plane for twist = $\pi$ (left) and twist = $0$ (right). 
The Higgs zeros are located at $(0,2)$ and $(0,-2)$, shown as filled and unfilled circles.
}
\label{gaugePi}
\end{figure}

To get an antimonopole, we can simply invert the ${\hat \phi}^a$ for the monopole.
This gives
\be
\phi^{a}= - h(r)\hat{r^{a}} =  \frac{h(r)}{r} (-x,-y,-z)
\label{invertedm}
\ee
However, this is not the only possibility. Any further local rotation of the directions of 
$\phi^a$ will also have the topology of an antimonopole. These local rotations are
irrelevant if we consider an antimonopole in isolation and all such gauge rotated
antimonopoles have the same energy. However, when we patch a monopole and 
an antimonopole together, there is an alignment issue, and the monopole-antimonopole
pair may have different energies depending on their ``relative twist''. For example,
in Fig.~\ref{gaugePi} we show Higgs vectors for the monopole described above and
the antimonopole configuration of Eq.~(\ref{invertedm}). This monopole-antimonopole
configuration has twist equal to $\pi$. In Fig.~\ref{gaugePi} we also show the zero twist
case, in which only the third component of the Higgs is inverted while the first and
second component are not
\be
\phi^{a}= \frac{h(r)}{r} (+x,+y,-z)
\label{invertedmzonly}
\ee

\begin{figure}
  \includegraphics[height=0.35\textwidth,angle=0]{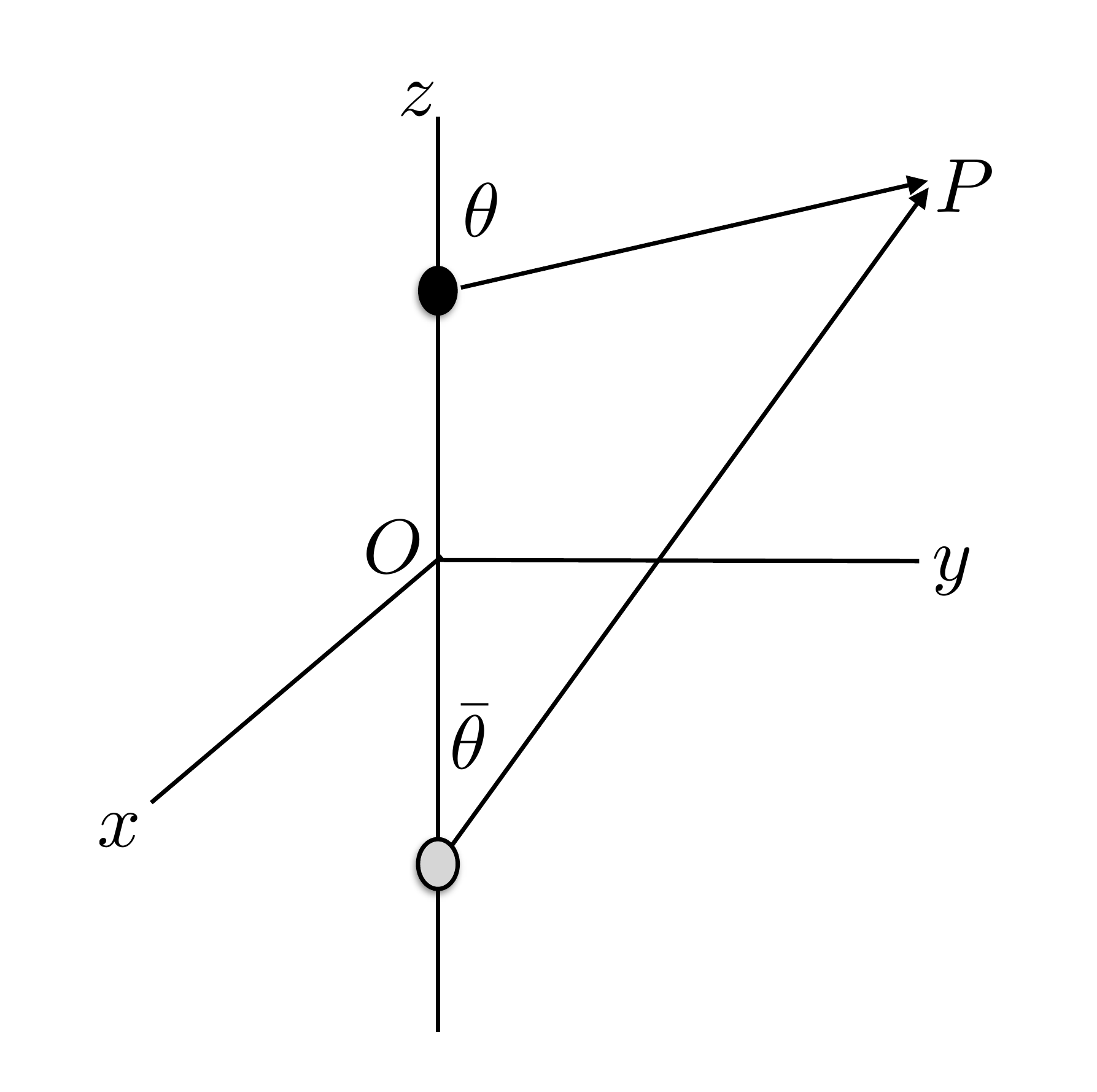}
  \caption{The physical configuration of the monopole-antimonopole pair.
 }
\label{dumbbell}
\end{figure}

Intermediate between the two cases of Eqs.~(\ref{invertedm}) and (\ref{invertedmzonly}), 
there is a continuous set of configurations that can be obtained by rotations of the scalar
field directions along the $z$-axis. 
The general configuration of the twisted monopole-antimonopole Higgs field can be written as
\ba
\hat{\phi}^{1}&=&(\sin\theta \cos\bar{\theta} \cos \gamma-\sin\bar{\theta}\cos\theta) \cos(\varphi-\gamma/2) \nonumber \\ 
&& \hskip 2.5 cm
 -\sin\theta \sin\gamma \sin(\varphi-\gamma/2)
 \label{phi1} \\
\hat{\phi}^{2}&=&(\sin\theta \cos\bar{\theta}\cos\gamma-\sin\bar{\theta}\cos\theta)\cos(\varphi-\gamma/2)\nonumber \\ 
&& \hskip 2.5 cm
-\sin\theta \sin\gamma \cos(\varphi-\gamma/2)
\label{phi2} \\
\hat{\phi}^{3}&=&\cos\theta \cos\bar{\theta}+\sin\theta \sin\bar{\theta}\cos\gamma
\label{phi3}
\ea
where, as shown in Fig.~\ref{dumbbell}, $\theta$ and $\bar{\theta}$ are the angles measured from the
the z-axis to the position vectors centered at the monopole and antimonopole, and
$\varphi$ is the azimuthal angle; $\gamma$ is the relative twist angle and takes values from $0$ to $2\pi$. 
In Cartesian coordinates we can write these position vectors as
\begin{equation}
	r_m = |{\bf x}-{\bf x}_m |, \, r_{\bar{m}}= |{\bf x}-{\bf x}_{\bar{m}}|
	\label{rm}
\end{equation}
where ${\bf x}_m = (0,0,z_0)$ and ${\bf x}_{\bar{m}} = (0,0,-z_0)$. 
Therefore, Eqns.~(\ref{phi1})-(\ref{phi3}) are expressed in Cartesian system as follows
\ba
	r_m r_{\bar{m}} \hat{\phi^1}& = & \left(cx+sy \right) \left[ (z+z_0) \cos \gamma -(z-z_0) \right]  \nonumber \\ &-& (cy-sx) r_{\bar{m}} \sin \gamma \\
	\label{r1}
	r_m r_{\bar{m}} \hat{\phi^2} & = & \left(cy-sx \right) \left[ (z+z_0) \cos \gamma -(z-z_0) \right]  \nonumber \\ &+& (cx+sy) r_{\bar{m}} \sin \gamma \\
	\label{r2}
r_m r_{\bar{m}} \hat{\phi^3} & =& (z-z_0)(z+z_0)+(x^2+y^2) \cos \gamma
	\label{r3}
\ea
where $c \equiv \cos \gamma$ and $s \equiv \sin \gamma$.
With this ansatz, we can write our initial guess for the Higgs field
configuration,
\be
\phi^{a}=h(r_{m})h(r_{\bar{m}})\hat{\phi}^{a}
\label{initialphi}
\ee
Our ansatz for the gauge fields follows from the requirement that the covariant derivatives
of the Higgs isovector vanish, $D_{\mu}\hat{\phi}=0$, at spatial infinity. This gives
\be
W_{\mu}^{a}=-\epsilon^{abc}\hat{\phi}^{b}\partial_{\mu}\hat{\phi}^{c}
\ee
We include profile functions to obtain our initial guess for the gauge fields,
\be
W_{\mu}^{a}=-(1-k(r_{m}))(1-k(r_{\bar{m}}))\epsilon^{abc}\hat{\phi}^{b}\partial_{\mu}\hat{\phi}^{c}
\label{initialW}
\ee
This initial guess automatically satisfies the asymptotic conditions in Eq.~(\ref{asymptotics})
for finite energy configurations. 

We can see that the twist has a gauge invariant meaning in two ways. First, the 
energy is gauge invariant and by explicit calculation we see that the energy of 
the configuration depends on the twist. Second, the twist can be expliclty 
defined in terms of the Chern-Simons number as discussed 
in~\cite{mantonklinkhamer}.  The bound state solution with
twist of $\pi$ is the sphaleron with Chern-Simons number of $1/2$.

\section{Twisted Dipole Gauge}
\label{twisteddipolegauge}

We would like to minimize the energy in Eq.~(\ref{energydensity}) but with the constraints that the monopole 
and antimonopole locations and their relative twist are held fixed. We have found a simple scheme
to impose such constraints, in part by making use of the topology of the monopole and antimonopole.
The key realization is that local gauge transformations can be made to freely choose the direction
${\hat \phi}^a$ at any spatial point. For example, the simplest choice would be to adopt the ``unitary
gauge'' in which the Higgs is spatially uniform. However, then the gauge fields are singular and this
makes the unitary gauge unsuitable for numerical work. Instead we adopt the ``twisted dipole'' gauge
which is that ${\hat \phi}^a$ is fixed by Eqs.~(\ref{phi1}), (\ref{phi2}), and (\ref{phi3}) throughout 
the numerical relaxation. This gauge choice automatically fixes the locations 
of the monopole and antimonopole due to the topology, and it also fixes the 
twist. The locations of the monopoles are chosen to lie within a cell of the 
lattice,
not on a vertex. This avoids evaluation of the fields at the centers of the monopoles and the
possibility of any fluctuations during field relaxation that can move the 
location of the monopoles.

Since we fix the direction of Higgs field isovectors at each spatial point, only the magnitude of the
Higgs field can vary and it is unnecessary to relax each of the components separately. Instead
we write $\phi^a=|\phi| \hat{\phi}^a$ and relax $|\phi|$ according to the equation
\ba
\nabla^2|{\phi} | &=& |{\phi} |\partial_i\hat{\phi}^a\partial_i\hat{\phi}^a  
+ g^2 |{\phi} | W_{i}^{a} W_{i}^{a} 
\nonumber \\ && \hskip 0 cm
- g^2 |{\phi} | W_{i}^{a} W_{i}^{b} \hat{\phi}^a \hat{\phi}^b  
- 2g |{\phi} | \epsilon^{abc}W_{i}^{a}\partial_i\hat{\phi}^b \hat{\phi}^c 
\nonumber \\ && \hskip 0 cm
+ \lambda (|{\phi} |^2-1)|{\phi} |
\label{magphieq}
\ea
Thus, we have 1 equation for $|{\phi}|$, 9 equations for $W^a_i$, and 3 equations for $\Gamma^a$.
However, in the static case, and since we are working in temporal gauge, the equations for
$\Gamma^a$ are trivial. This leaves us with 10 non-trivial equations to solve.

\section{Numerical Solution}
\label{numerics}

To see how our numerical scheme works, we first set all the time derivatives to zero in the equations 
for $|{\phi}|$ and $W^a_i$ and discretize the spatial derivatives. Our discretized equations at a given
lattice point can be written in the following generic form
\be
{\bf E}[\{f_\beta \}] = 0
\label{genericequation}
\ee
where $\{f_\beta \}$ denotes the set of fields, and ${\bf E}$ is the array of discretized
equations obtained from Eqs.~(\ref{eqphi})-(\ref{Gammaeq}).
Now, if we use second order spatial derivatives, the Laplacian term 
in these equations can be written as 
\begin{alignat}{2}
\nabla^2 f(i,j,k) &\to -\frac{6}{\delta^2} f(i,j,k) 
+ \frac{1}{\delta^2} [ f(i+1,j,k)\nonumber \\ &+ f(i-1,j,k) + f(i,j+1,k)
+ f(i,j-1,k) \nonumber \\ &+ f(i,j,k+1)+f(i,j,k-1)] \nonumber
\label{secondorder}
\end{alignat}
where $f$ denotes any one of the fields and $\delta$ is the lattice spacing. Then we re-write Eq.~(\ref{genericequation})
for the field $f_\alpha$ as
\be
 f_\alpha (i,j,k) = \frac{\delta^2}{6} {\bf E}_\alpha [\{f_\beta \}] + f_\alpha(i,j,k) 
\ee
So far this is exactly equivalent to Eq.~(\ref{genericequation}), but now we take the left-hand side at the
current ($n^{\rm th}$) iteration step and the right-hand side at the previous iteration step
\begin{alignat}{2}
 f_\alpha^{(n)} (i,j,k) = \frac{\delta^2}{6} {\bf E}_\alpha [\{f_\beta^{(n-1)} \}] + f_\alpha^{(n-1)} (i,j,k)
\label{iterationequation}
\end{alignat}
In fact, once a field is updated at some point $(i,j,k)$, that value is immediately used on the right-hand side
for the next computation.
In our numerical runs, we employ this approach but use sixth order derivatives for better accuracy.
Then the numerical coefficient of the ${\bf E}_\alpha$ term is $6/49$ instead of $1/6$. 

For most of our simulations, we chose a cubic lattice with $128^{3}$ lattice points
and lattice spacing $\delta =0.2$ and Dirichlet boundary conditions. Since we have 
set $g=1=\eta$, the mass of the heavy gauge fields is $m_v=g\eta = 1$ and the
scalar mass is $m_s=\sqrt{2\lambda}\eta =\sqrt{2\lambda}$. The monopole
width is primarily set by the mass of the vector field and so the core of the
monopole is resolved by $\sim 5^3$ lattice points in our simulations.
 
The monopole and antimonopole locations are fixed at $z=\pm (z_0 + \delta /2)$
respectively. With the offset by half a lattice spacing, we ensure that the zeros
of the Higgs field do not lie at a lattice point and there are no artificial numerical
singularities due to $1/r$ factors when specifying initial conditions as in 
Eqs.~(\ref{initialphi}) and (\ref{initialW}).
 
We perform runs with different values of the coupling constant $\lambda$, twist $\gamma$, 
and monopole-antimonopole separation $d =2 z_0$. We ran our code for each
set of parameters for 1000 iterations and then found the asymptotic value of energy by 
extrapolating the energy vs. iteration number power law dependence to infinite number
of iterations.

\begin{figure}
\subfloat{\includegraphics[trim={2.6cm 1.0cm 0 0},height=0.23\textwidth,angle=0]{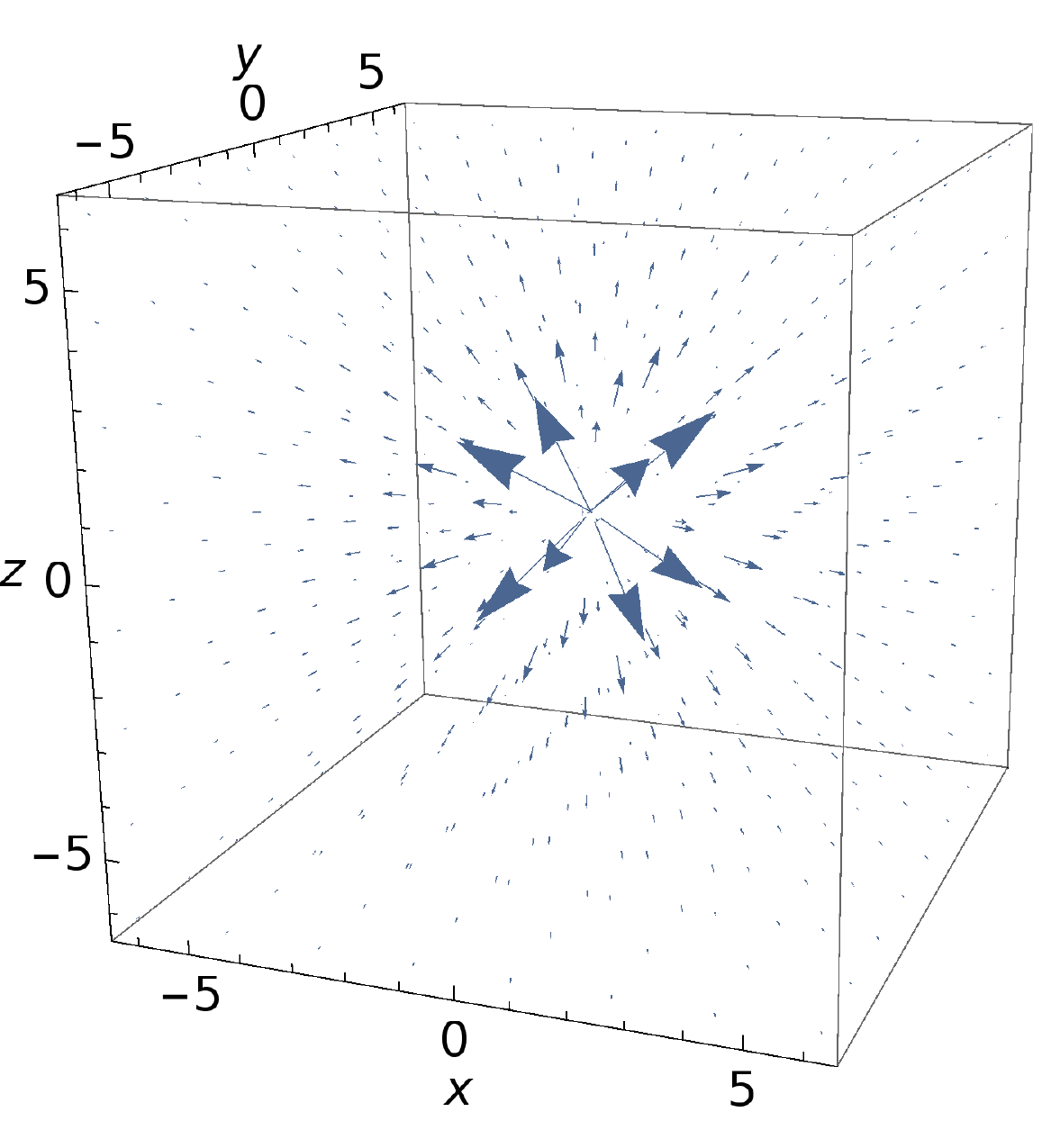}} 
\subfloat{\includegraphics[trim={0.2cm 0 2.5cm 1.6cm},height=0.19\textwidth,width=0.24\textwidth,,angle=0]{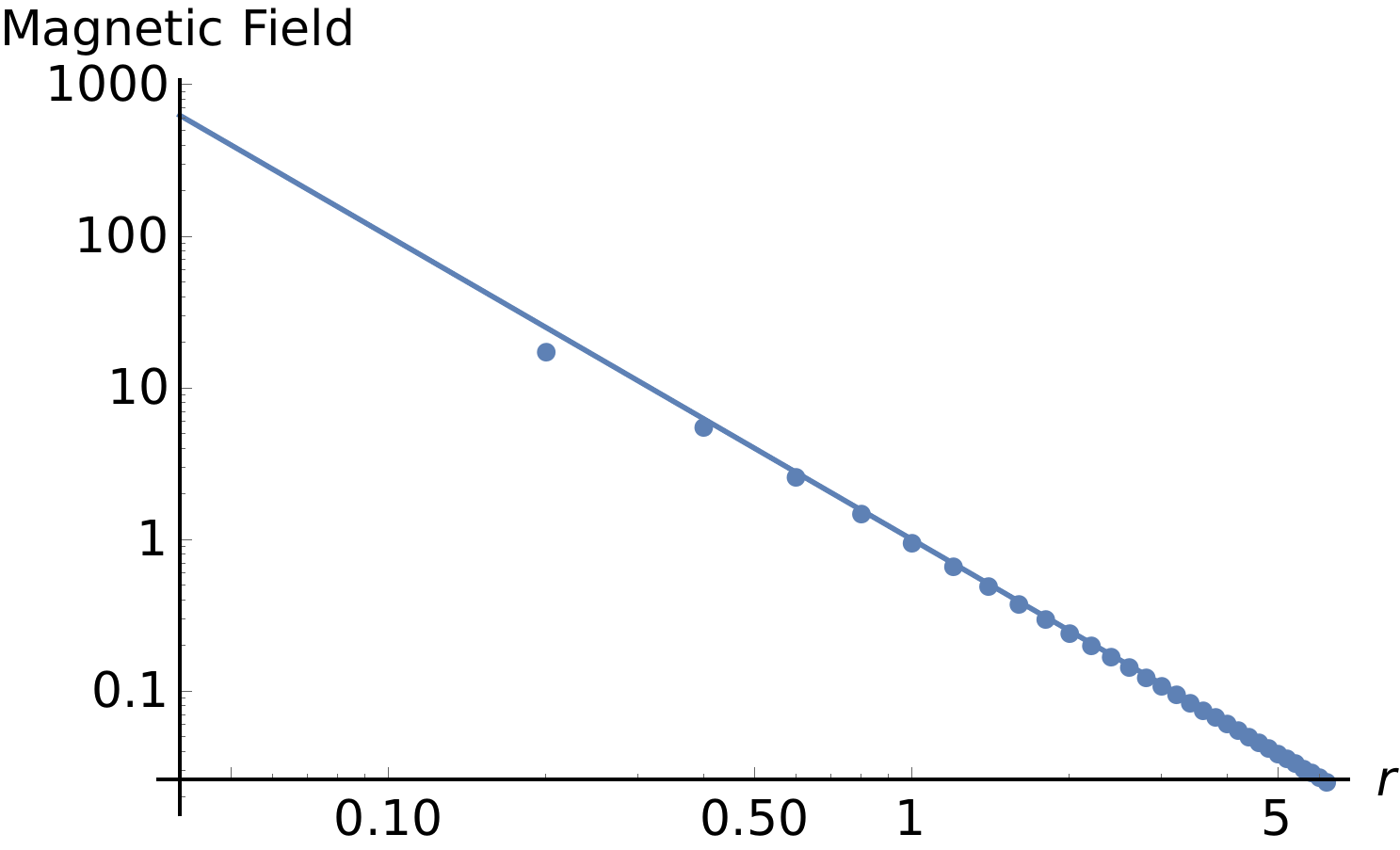}}
\caption{A 3D vector plot of the magnetic field of a single monopole.
The log-log plot of magnetic field strength of the monopole vs distance $r$ for $\lambda=4$. 
The dots represent the numerical solution and the solid line shows a $1/r^2$ fit. 
%\TV{figure placement can be improved.}
}
\label{singlemonopole}
\end{figure}

We validated our numerical scheme through various means. First, we solve the equations for a 
single monopole with coupling parameter, $\lambda=4$, and in the hedgehog gauge on a $64^3$ 
lattice. The magnetic field from this solution is found to precisely fall as $\propto r^{-2}$ away from 
the location of Higgs zero as shown in the Fig.~\ref{singlemonopole}. 
Second, for each set of parameters $\lambda$ and $\gamma$, the
energy for the monopole-antimonopole asymptotes to twice the monopole mass at large separation.
Third, we find that the energy has a saddle point at twist=$\pi$ for all values of $\lambda$ 
that we have considered. This is consistent with the general arguments by Taubes~\cite{taubes}
and his analysis for the $\lambda=0$ case.

\begin{figure}
\includegraphics[height=0.24\textwidth,angle=0]{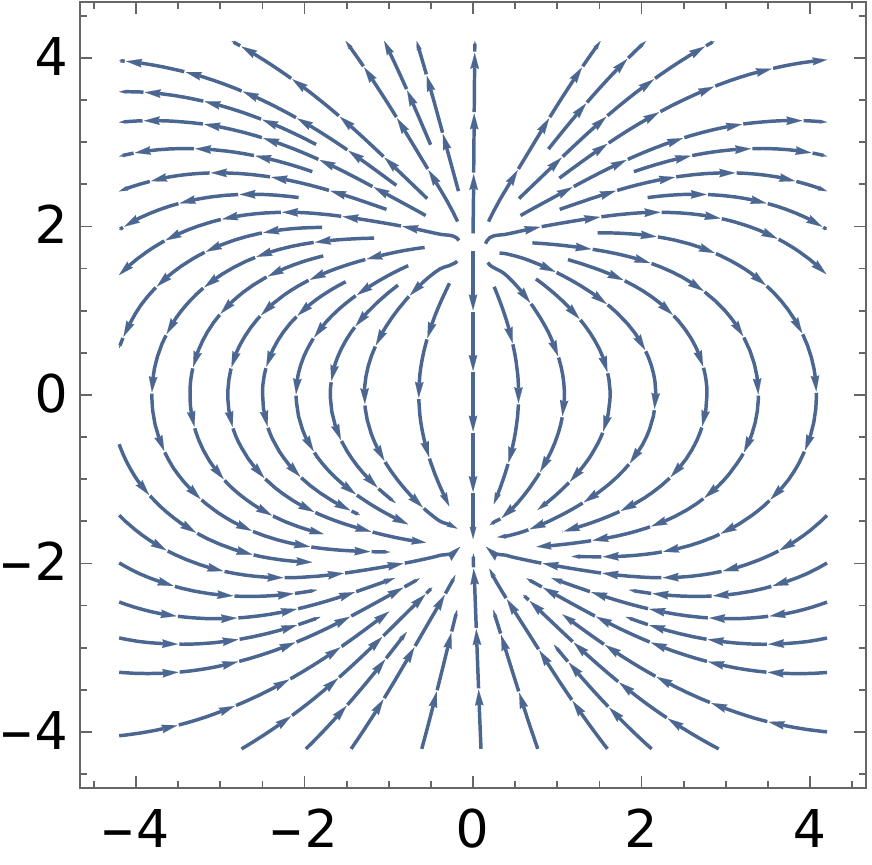}
\includegraphics[height=0.24\textwidth,angle=0]{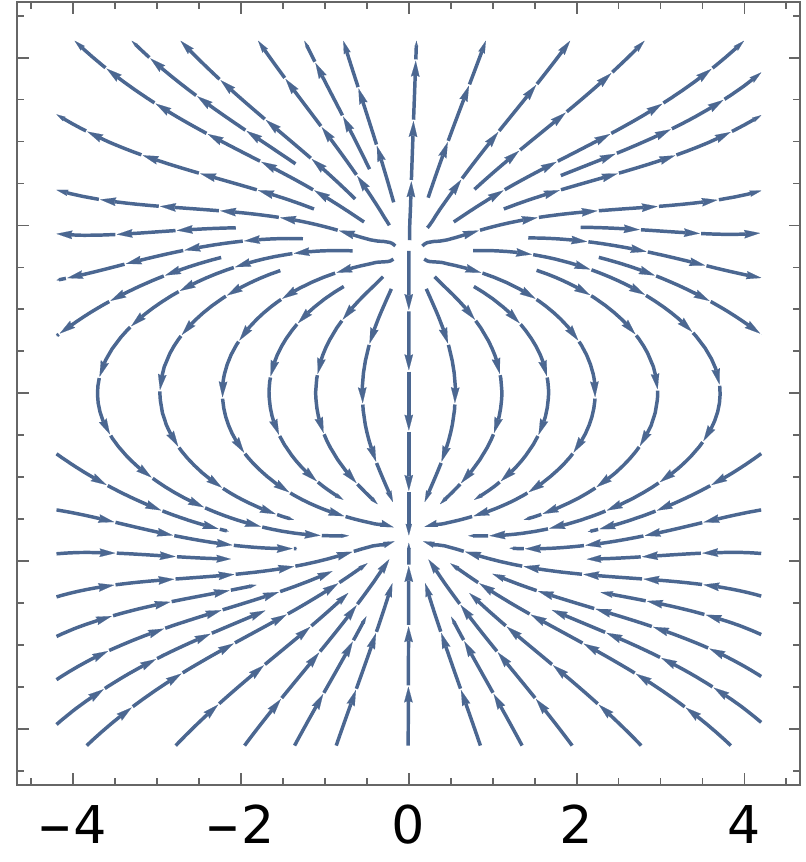}
\caption{Magnetic field lines for $\lambda=4$, $d=3.4 \, (z_0 = 1.7)$ in the $xz$-plane in
  the untwisted case (left) and the maximally twisted case (right).% \TV{fix figure.}
 }
\label{mflines}
\end{figure}

\section{Results and Conclusions}
\label{results}

We start with results for the magnetic field lines for a monopole-antimonopole pair with and without
twist. The results are shown in Fig.~\ref{mflines}. For the untwisted case and for small separations, 
when the boundary effects are not significant, we have checked that the magnetic field strength falls 
off as $r^{-3}$ within our lattice, just as we would expect for a magnetic dipole.

\begin{figure}
  \includegraphics[height=0.25\textwidth,angle=0]{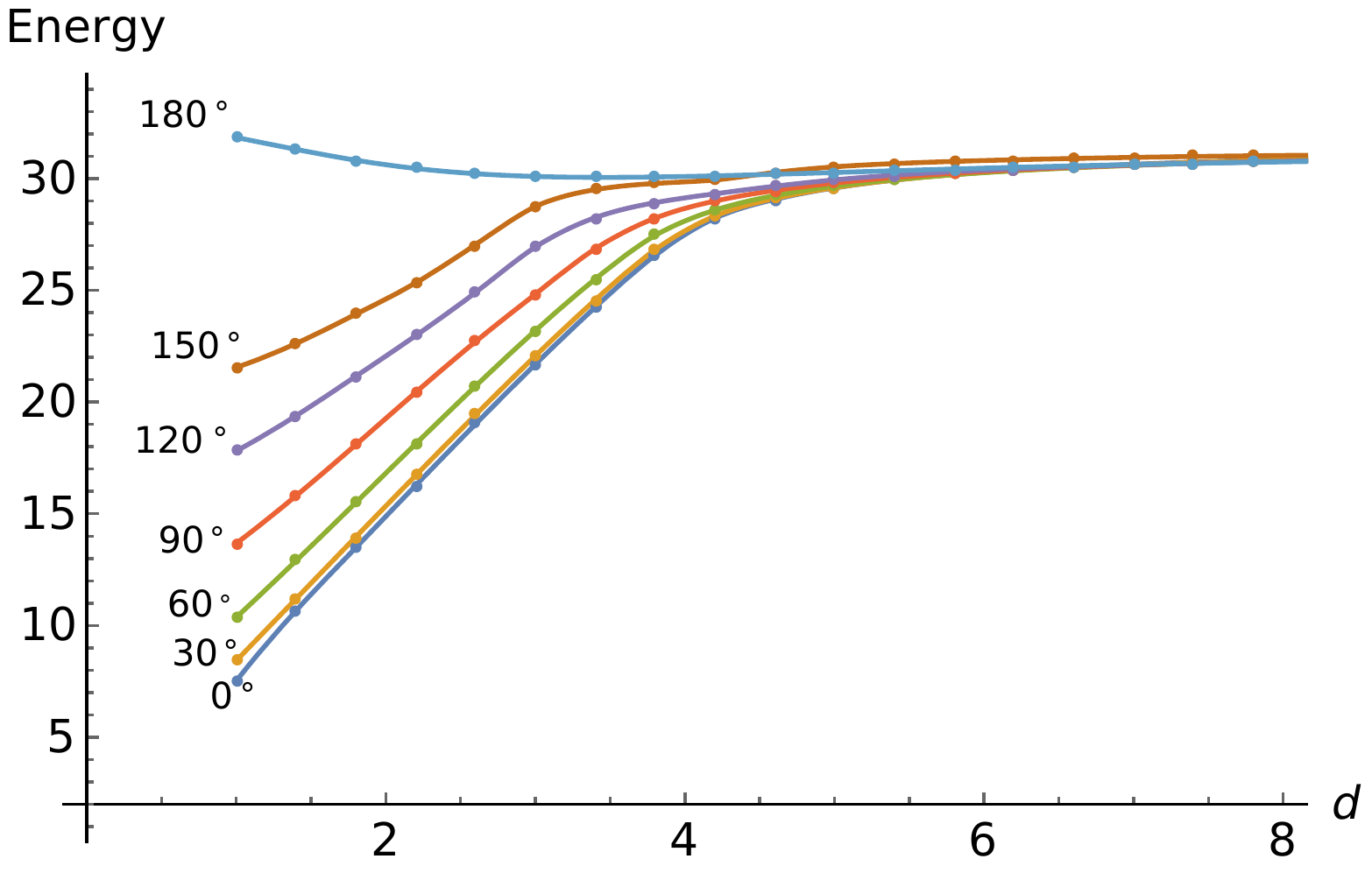}
  \caption{Total energy as a function of monopole-antimonopole separation $d$ 
  for $\lambda=1$ and twist varying from $0$ to $\pi$.
 }
\label{lambda1}
\end{figure}

\begin{figure}
  \includegraphics[height=0.25\textwidth,angle=0]{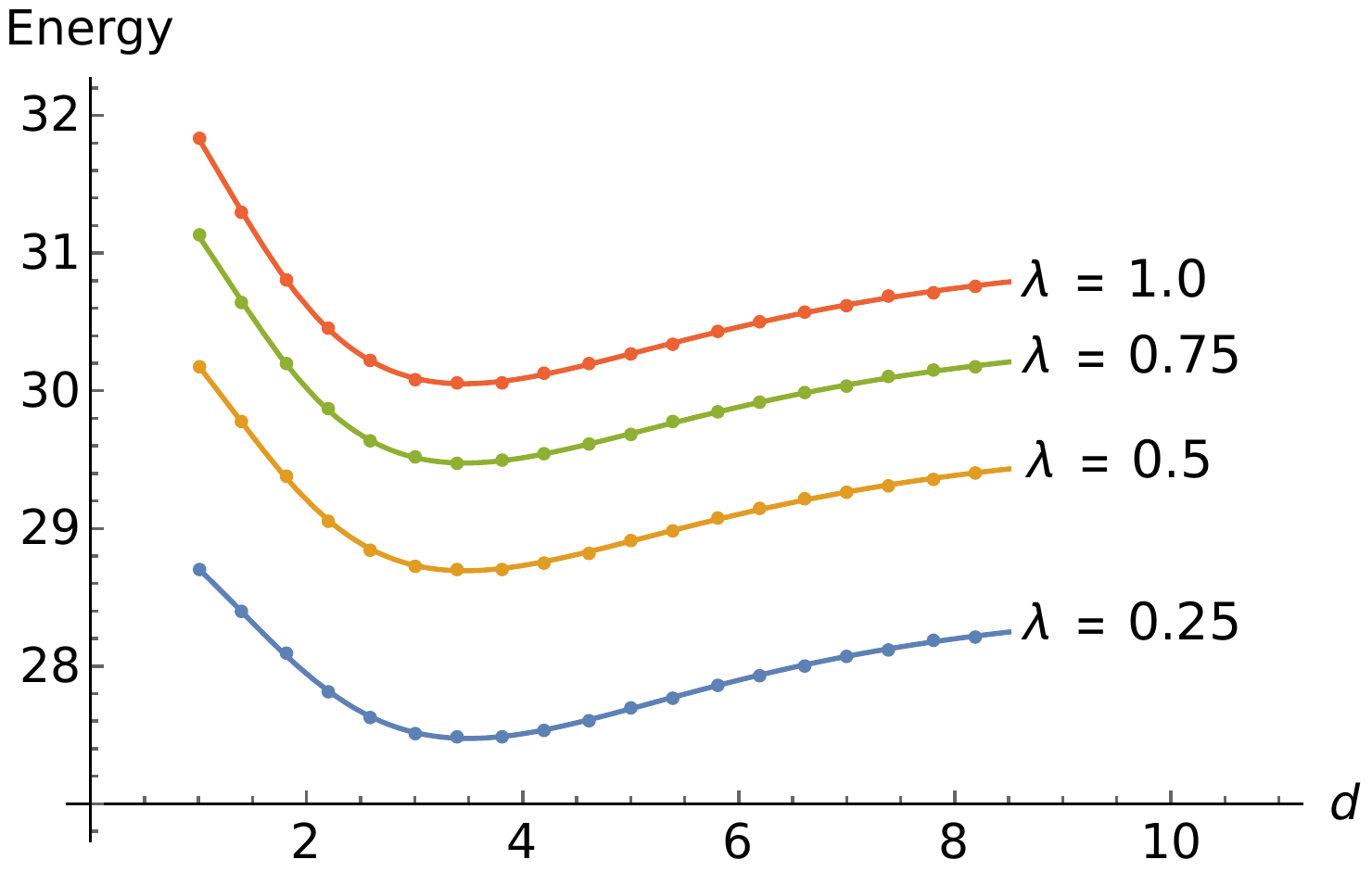}
  \caption{Total energy as a function of monopole-antimonopole separation $d$ for twist 
  $\gamma = \pi$ and $\lambda$ varying from $0.25$ to $1.0$. The sphaleron solution 
  is at the minimum in every curve.}
\label{twist180}
\end{figure}

In Fig.~\ref{lambda1} we show the relaxed energy of the monopole-antimonopole vs. separation
for $\lambda=1$ and for several different twist values. At large separation, the total energy goes
to twice the monopole mass, as we expect since the Coulombic interaction dies off. At small
separations, the interaction is attractive for small values of twist and repulsive for very large values 
of  twist. The curve for $\gamma=\pi$ (maximum twist) has a minimum at $d \approx 3.4$. This is 
seen more clearly in Fig.~\ref{twist180} where we plot the relaxed energy vs.  
separation for $\gamma=\pi$ and for several different values of $\lambda$. A 
three-dimensional plot of energy vs.  separation and twist would have a saddle 
point in which
the minimum is along the direction of separation and a maximum along the twist direction.
This saddle-point solution which corresponds to a bound state of a monopole and antimonopole
is called a ``sphaleron'' ~\cite{mantonklinkhamer} and plays an important role in baryon number 
violating processes in particle physics. 

%\textbf{The curves in Fig.~\ref{lambda1} clearly seem to respect the schematics 
%behind the expression (Eq. \ref{taubesequation}) proposed by Taubes, but the 
%differences arise due to the fact that monopoles are not point particles and 
%annihilate as the separation between them becomes smaller, as can be seen in 
%Fig.~\ref{taubescurves}. To find an analytical expression for the curves that 
%fit our data, we use the following ansatz,}

\begin{figure}
  \includegraphics[height=0.25\textwidth,angle=0]{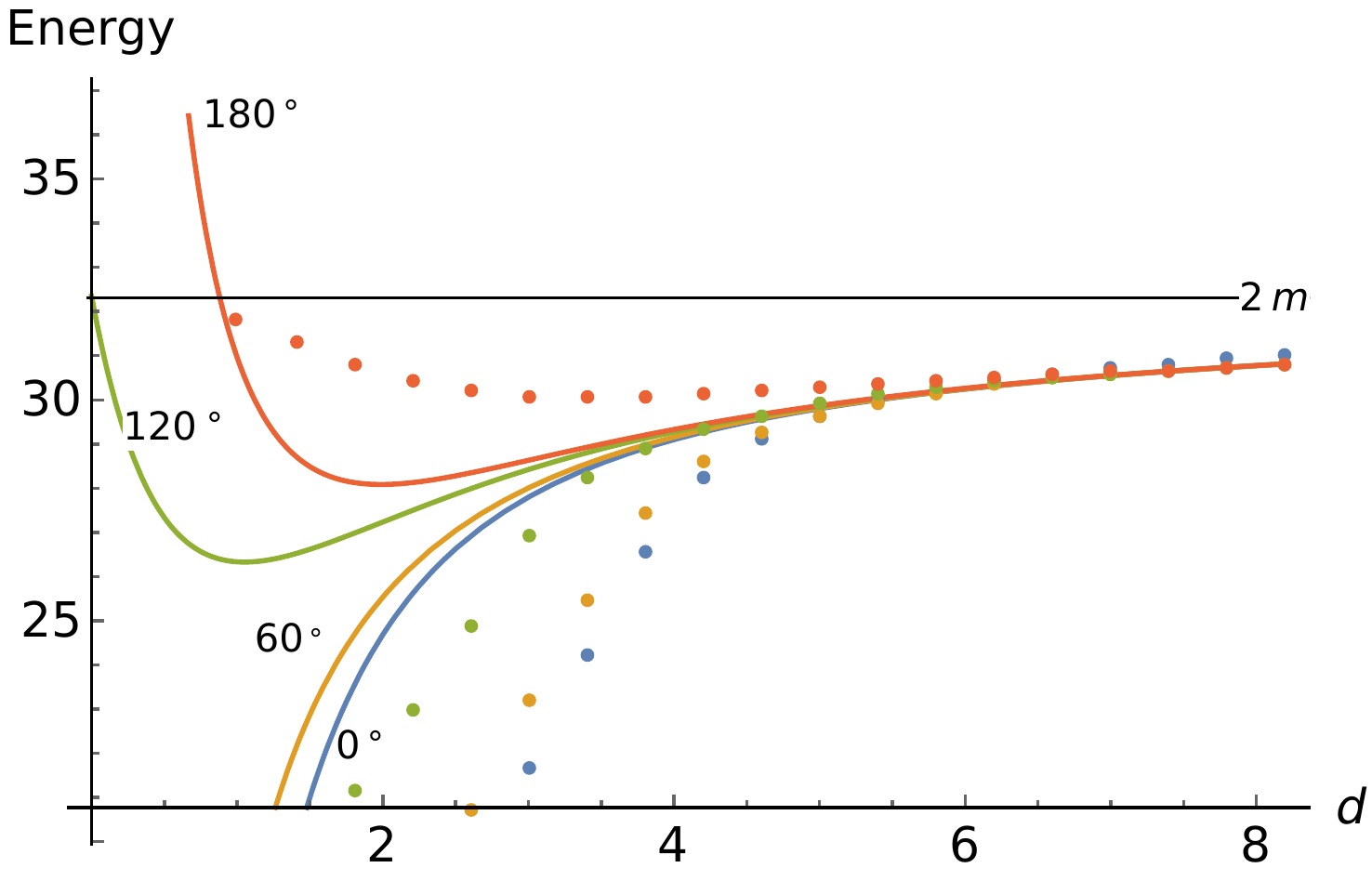}
  \caption{
%  Interaction energy as a function of monopole-antimonopole separation 
%	$d$ for $\lambda=1$ and twist varying from $0$ to $\pi$. Solid curves 
%	represent the function proposed by Taubes and dots represent our 
%	numerical data.  
Comparison of the data for $\lambda=1$ and the expression in
Eq.~(\ref{taubesequation}) plus twice the monopole mass (solid curves), 
demonstrating that the expression is not a good fit to the data.
	}
\label{taubescurves}
\end{figure}

The curves in Fig.~\ref{lambda1} have qualitative features of $V(d,\gamma)$ in
Eq.~(\ref{taubesequation}) but quantitative differences are apparent when
we overlay the analytic expressions and the data as shown in Fig.~\ref{taubescurves}.
As discussed in the introduction, the differences arise since monopoles are not point 
particles and monopole-antimonopole can partially annihilate as the separation 
between them becomes smaller. This annihilation leads to vanishing total energy 
as the separation goes to zero in the untwisted case unlike the divergent energy 
predicted by Taubes' potential.  

To quantify the energy reduction due to annihilation we write
\ba
E_{\rm data}(d, \gamma) &=& A(d,\gamma) E_{\rm Taubes} (d,\gamma) \nonumber \\
&=& A(d,\gamma ) \left [ 2m + V(d,\gamma) \right ]
\label{Adefn}
\ea
where $E_{\rm data}$ is the energy of the monopole-antimonopole with separation 
$d$ and twist $\gamma$ as computed numerically,
$m$ is the mass of a single monopole,
$2m+V(d,\gamma)$ is the energy as determined using the Taubes formula in
Eq.~(\ref{taubesequation}) valid for point-like monopoles, and $A(d,\gamma)$ is 
an energy-reduction factor arising due to the
finite core size of the monopoles. At large separations $A(d,\gamma)$
goes to one because then the point-like approximation is valid.

\begin{figure}
  \includegraphics[height=0.25\textwidth,angle=0]{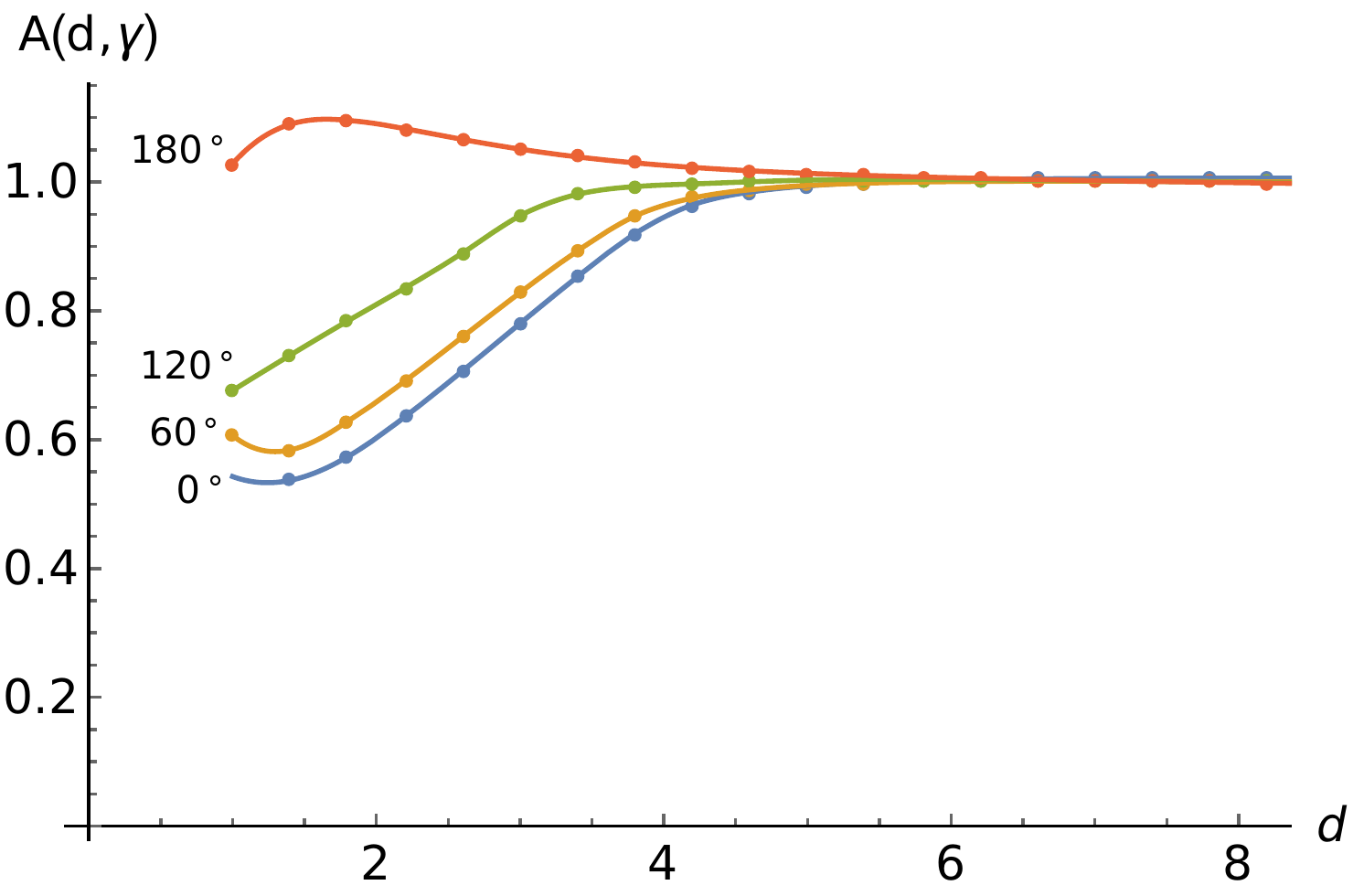}
  \caption{The ``annihilation'' function $A(d,\gamma)$ defined in 
	Eq.~(\ref{Adefn2}) vs. $d$ for some values of $\gamma$.
  }
\label{Aplots}
\end{figure}

We use Eq.~(\ref{Adefn}) to determine $A$ as
\be
A(d,\gamma) = \frac{E_{\rm data}(d, \gamma)}{E_{\rm Taubes}(d,\gamma)}
\label{Adefn2}
\ee
and we plot $A(d,\gamma)$ for several values of $\gamma$ in Fig.~\ref{Aplots}.  
These plots quantify the partial annihilation of monopole
and antimonopole due to their finite core sizes. As expected, $A \to 1$
at large separation because the point-like approximation gets better.
At small separation, the computed energy is smaller 
than the energy predicted from the Taubes formula due to partial
annihilation. From the curves for different $\gamma$ values, we see that the 
annihilation is less effective as the twist increases. This too is expected 
because annihilation can only occur if the fields are aligned in suitable ways 
while the twist forces them to be misaligned (see Fig.~\ref{gaugePi}). In our 
plot
we see that the maximally twisted case has $A$ that is $\sim 10\%$
greater than 1 at short distances. We think this is due to small numerical errors 
or small corrections to $E_{\rm Taubes}$ that have not been taken into account.

The qualitative behavior of $A(d,\gamma)$ can be written as
\be
A(d,\gamma) \sim \tanh \left ( \frac{d}{1+\cos\gamma} \right ).
\ee

%To find an analytical fit to our data, we 
%\TV{have experimented with different functional forms. We have found a 
%reasonable fit with the following ansatz,}
%%use the following ansatz,
%\be V(d) = 4\pi(1-e^{-d})\left(-\frac{1}{d} -\frac{2Be^{-d}}{d^A} - \frac{e^{-\sqrt{2\lambda}d}}{d}\right).
%\label{fitfunction}
%\ee
%\TV{
%The overall factor of $1-e^{-d}$ is included so that the potential does not diverge 
%at small separations. When the monopole-antimonopole separation becomes less
%than the core size, the cores overlap and we expect a reduction in the energy.
%%We find the reasonably good 
%The fit to the data is shown in Fig.~\ref{energyfit} for $\lambda=1$
%with $A \approx -0.5$; $B$ a parameter that varies with twist as shown in 
%Fig.~\ref{Bfit}. We have not found a good explanation for this value of $A$ and
%the behavior of $B$.
%}
%%
%%With $A \approx -0.5$ and $B$ a parameter that varies with twist as shown in 
%%Fig.~\ref{Bfit},
%%we get the reasonably good fit shown in Fig. \ref{energyfit} for $\lambda=1$.

%\begin{figure}
%  \includegraphics[height=0.25\textwidth,angle=0]{Bfit.pdf}
%\caption{
%%	The curve fit parameter $B$ in Eq. \ref{fitfunction} as a 
%%	function of the twist angle $\gamma$. 
%The curve fit parameter $B$ in Eq.~(\ref{fitfunction}) as a function of the 
%	twist angle $\gamma$.
%}
%\label{Bfit}
%\end{figure}

\begin{figure}
	\subfloat{\includegraphics[trim={2.6cm 0 0 0},height=0.26\textwidth,angle=0]{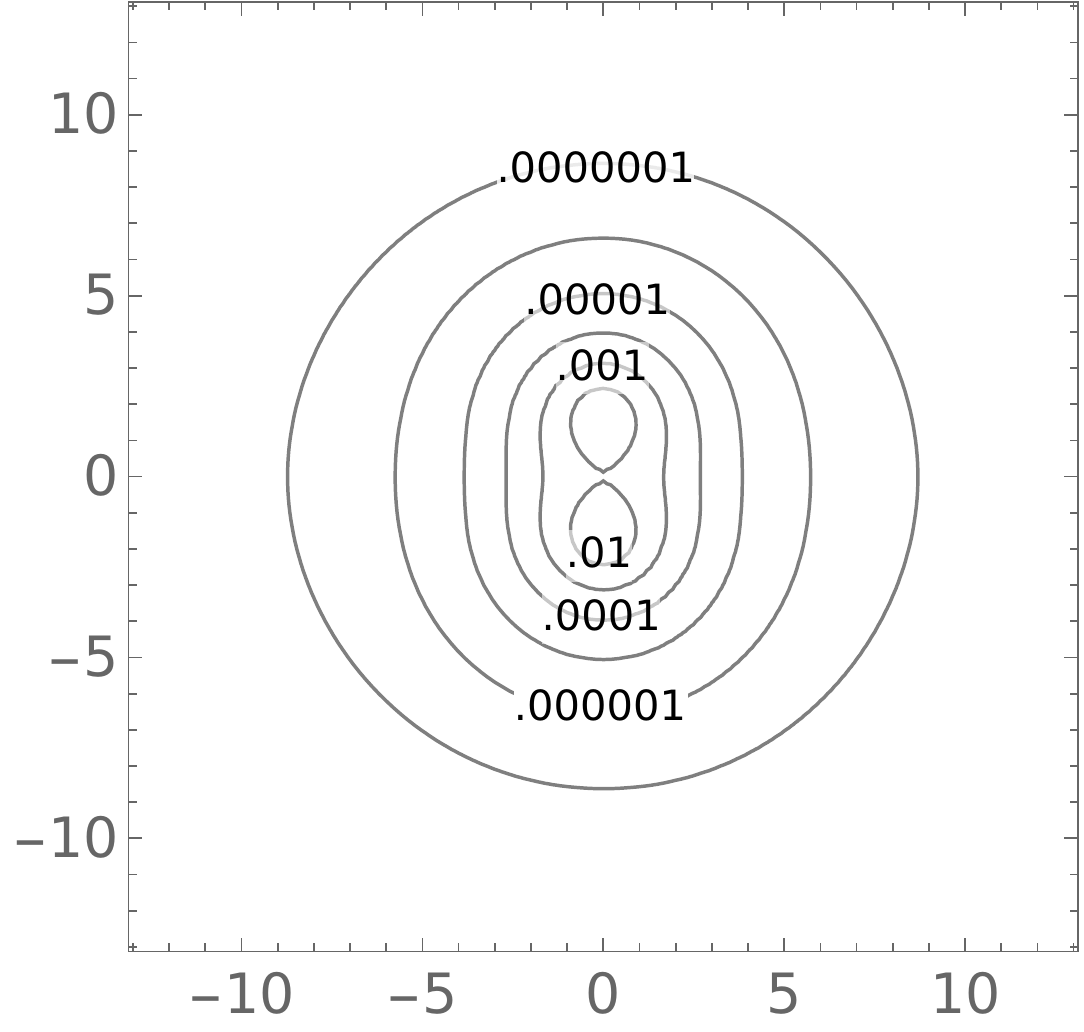}} \qquad
\subfloat{\includegraphics[trim={1.8cm 0 1.8cm 0},height=0.26\textwidth,angle=0]{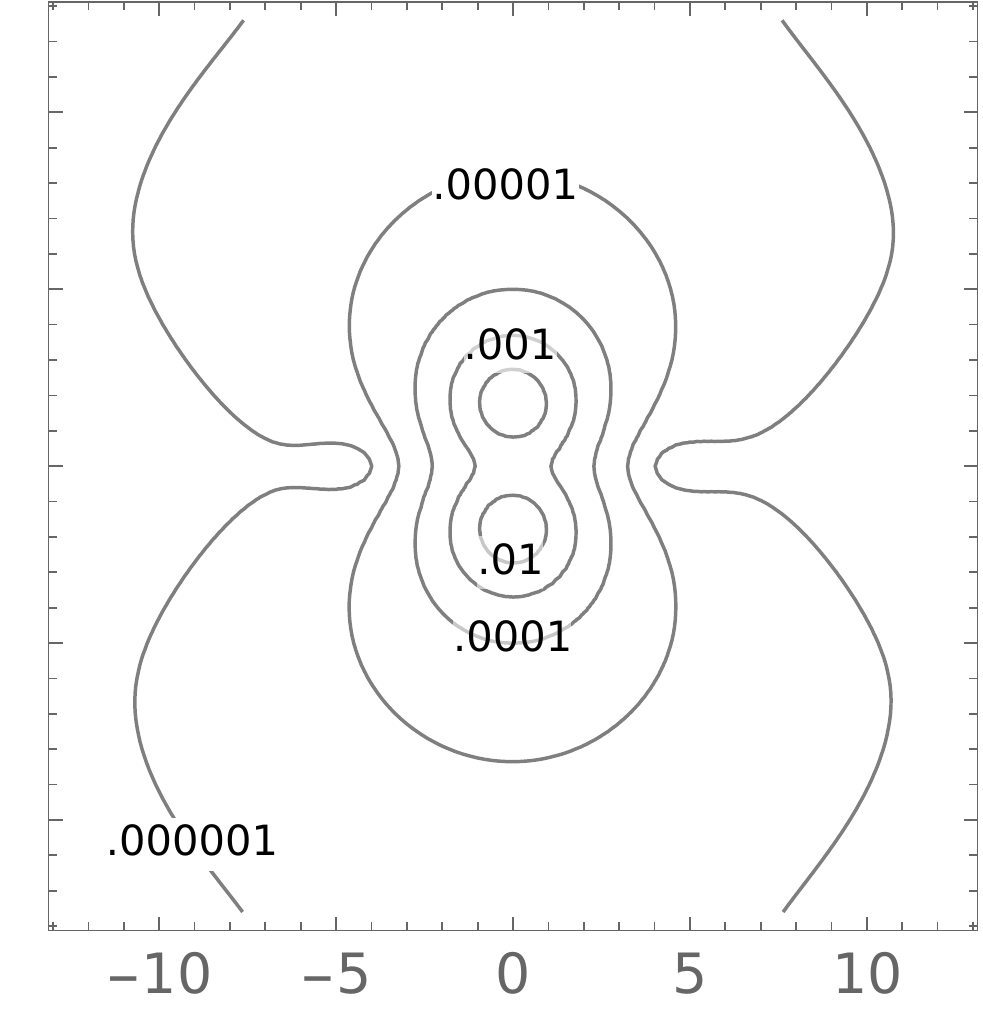}}
  \caption{Energy density contours for $\lambda=4$, $d=3.4$ in the $xz$-plane in
  the untwisted case (left) and the maximally twisted case (right).% \TV{fix figure.}
 }
\label{sphaleroncontours}
\end{figure}

\begin{figure}
  \includegraphics[height=0.25\textwidth,angle=0]{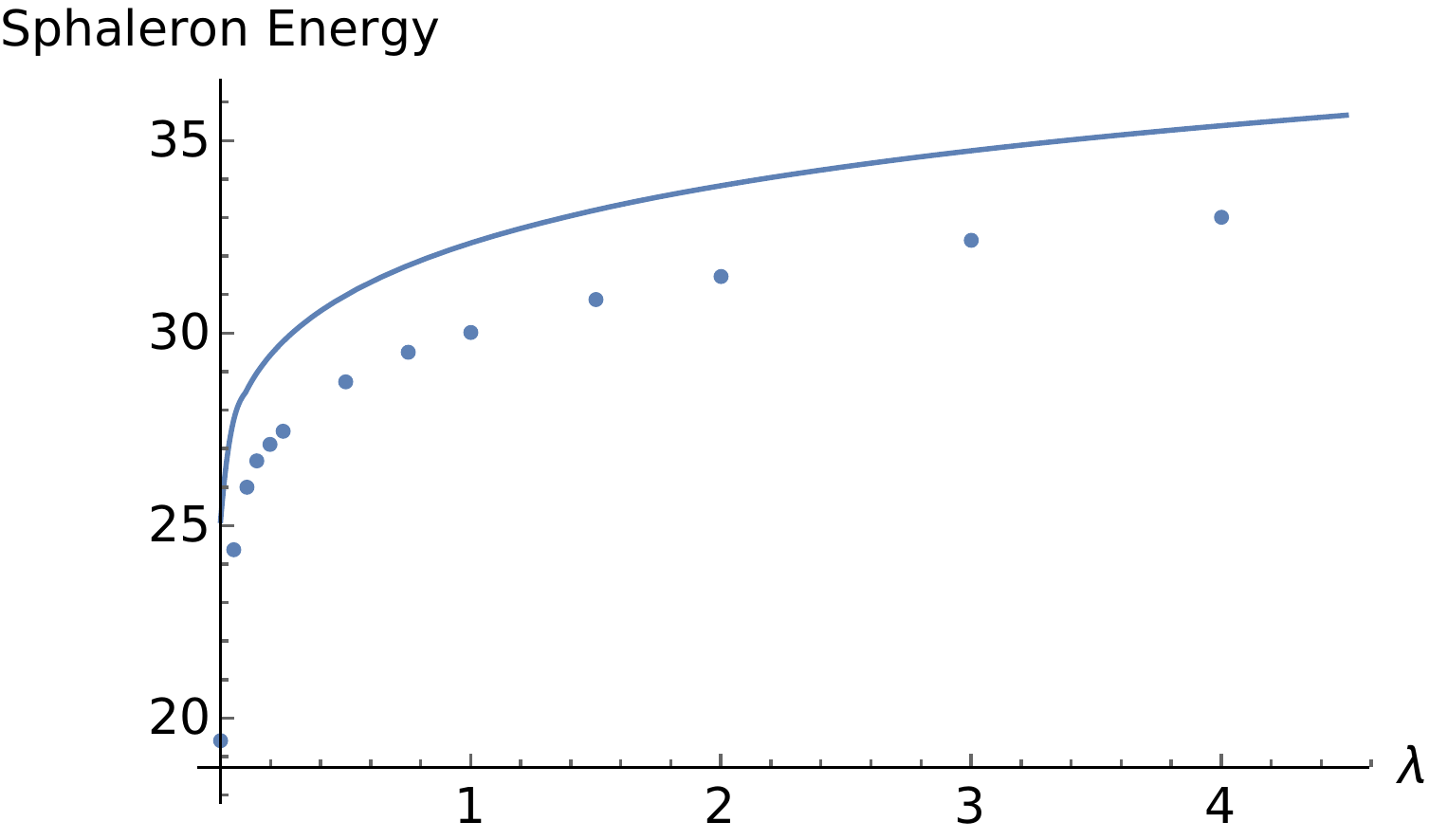}
  \caption{Sphaleron energy as a function of $\lambda$ (dots). The solid curve shows twice
  the monopole mass vs. $\lambda$.
 } 
\label{energyvslambda}
\end{figure}

\begin{figure}
  \includegraphics[height=0.26\textwidth,angle=0]{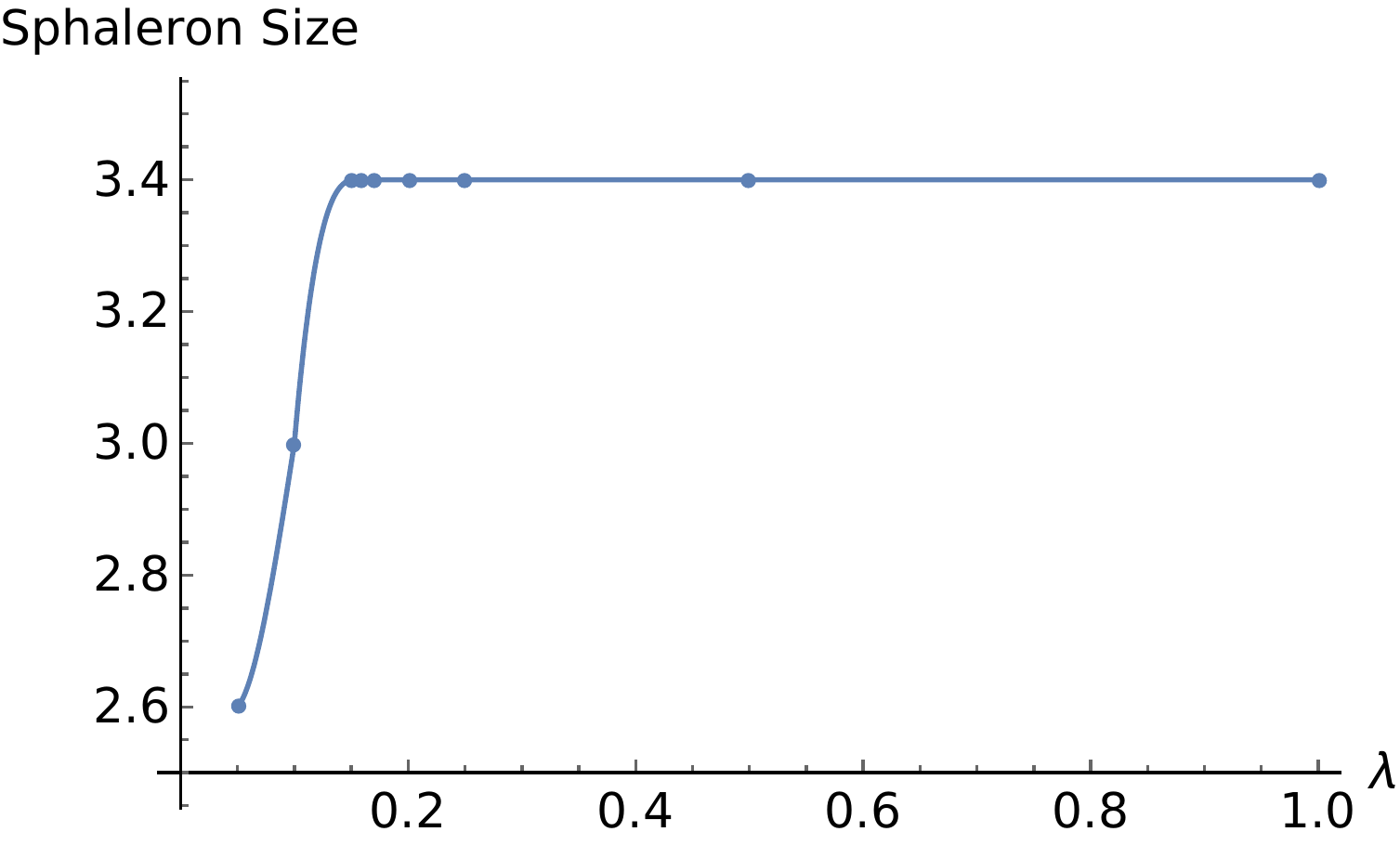}
  \caption{Monopole-antimonopole separation in the sphaleron, $d_s$, vs. $\lambda$. 
 } 
\label{separationvslambda}
\end{figure}

In Fig.~\ref{sphaleroncontours} we show energy contours of the untwisted monopole-antimonopole
pair and also the sphaleron solution. The total energy of the sphaleron, $E_s$, 
depends on the coupling constant $\lambda$ as shown in 
Fig.~\ref{energyvslambda}. The monopole-antimonopole separation within the 
sphaleron solution, $d_s$, depends weakly
on $\lambda$ for large values of $\lambda$ as can be seen in Fig.~\ref{separationvslambda}.
Since some fields fall off very slowly as $\lambda \rightarrow 0$, our 
predicted total energy at such small values of coupling constant could be 
underestimates by at most $20 \%$ (we predict this error by comparing the 
numerically obtained mass of BPS monopole with the theoretical value of 
$4\pi$).  

To conclude, we have numerically constructed twisted monopole-antimonopole pairs and mapped out
their interaction energy for a range of coupling constants. We have explicitly 
confirmed the arguments made by Taubes~\cite{taubes} on the existence of a 
bound state solution of monopole and antimonopole, also called a sphaleron.  In 
addition, we have studied the dependence of the sphaleron energy and size on 
coupling constant.

Our results are significant also because they provide a method that can be used to accurately set up initial configurations for dynamical studies such as monopole-antimonopole scattering. In the electroweak context, the method can be used to set up electroweak dumbbell configurations ~\cite{nambu}.

\section{Acknowledgment}
AS thanks the MCFP, University of Maryland for hospitality. We also thank Erich 
Poppitz for useful comments. The computations were done on the A2C2 Saguaro 
Cluster at ASU.
This work is supported by the U.S. Department of Energy, Office of High Energy 
Physics, under Award No.~\uppercase{DE-SC0018330} at ASU.

\newpage
\bibliography{main}

\end{document}